\documentclass[12pt]{article}

\usepackage{amsmath}
\usepackage{amssymb}
\usepackage{amsfonts}
\usepackage{latexsym}
\usepackage{color}

\catcode `\@=11 \@addtoreset{equation}{section}

\catcode `\@=12

\newcommand{\be}{\begin{equation}}
	\newcommand{\en}{\end{equation}}
\newcommand{\bea}{\begin{eqnarray}}
	\newcommand{\ena}{\end{eqnarray}}
\newcommand{\beano}{\begin{eqnarray*}}
	\newcommand{\enano}{\end{eqnarray*}}
\newcommand{\bee}{\begin{enumerate}}
	\newcommand{\ene}{\end{enumerate}}

\newcommand{\HH}{\mathfrak H}
\newcommand{\R}{\mathbb{R}}
\newcommand{\mc}{\mathcal}

\newcommand{\D}{{\mc D}}

\newcommand{\Sc}{{\cal S}}

\newcommand{\F}{{\cal F}}
\newcommand{\G}{{\cal G}}

\newcommand{\J}{\mc J}
\newcommand{\Lc}{{\cal L}}

\newcommand{\scr}{{\Sc(\mathbb{R})}}

\newcommand{\1}{1 \!\! 1}

\newcommand{\Hil}{\mc H}

\newtheorem{thm}{Theorem}

\newtheorem{defn}[thm]{Definition}

\catcode `\@=11 \@addtoreset{equation}{section}
\catcode `\@=12

\begin{document}

\thispagestyle{empty}

\vspace*{2cm}

\begin{center}
{\Large \bf A pseudo-bosonic Klein-Gordon field with finite two-points function}   \vspace{2cm}\\

{\large F. Bagarello}\\
Dipartimento di Ingegneria,
Universit\`a di Palermo,\\ I-90128  Palermo, Italy\\
and I.N.F.N., Sezione di Catania\\
e-mail: fabio.bagarello@unipa.it\\

\end{center}

\vspace*{1cm}

\begin{abstract}

We introduce a class of pseudo-bosonic Klein-Gordon fields in 1+1 dimensions and we discuss some of their properties. This work  originates from non Hermitian quantum mechanics and  deformed canonical commutation relations. We show that, within this class of fields, there exist a specific subclass with the interesting feature of having finite equal space-time two-points function, contrarily to what happens for {\em standard} Klein-Gordon fields. This, in our opinion, is a relevant aspect of our proposal which is a good motivation to undertake a deeper analysis of this (and related) quantum fields.

\end{abstract}

\vspace{2cm}


\vfill


\newpage

\section{Introduction}

Quantum field theory (QFT) is an extremely interesting area of research, linked to many crucial aspects of physics, from elementary particles to many body systems and statistical mechanics. There are thousands of papers and monographs dealing with QFT, in its various aspects. We only cite here \cite{wig}-\cite{muta}, where more references can be found. What is particularly interesting for us in this paper is the fact that, see \cite{wig}, a quantum field is necessarily an operator valued distribution and, as such, taking its powers, or multiplying two such fields, is a dangerous operation: distributions cannot be multiplied, in general, even at a classical level. For instance $\delta(x)^2$ is something whose meaning is not clear. Quite often, in the literature, one finds claims stating simply that $\delta(x)^2$ does not exist. However, several attempts have been proposed in the past to define, in some rigorous way, this and other products of distributions. Results in this direction can be found in \cite{col}, or in some specific case in \cite{breme}-\cite{bagdist3}. In particular, in \cite{bagdist2}, we have proposed a regularization of a Klein-Gordon field (KGF) $\varphi(x,t)$ in the attempt to produce finite equal-time two points function $\Delta_+(x,y)=\langle \Psi_0, \varphi(x,t) \varphi(y,t) \Psi_0\rangle$ in the limit $y \rightarrow x$. Here $\Psi_0$ is the vacuum of the bosonic operators used in the usual expansion of the field  $\varphi(x,t)$, \cite{itz,bd}. However, the approach proposed in \cite{bagdist2} does not work, in the sense that the limit of $\Delta_+(x,y)$ for $y\rightarrow x$, even after the regularization proposed therein, diverges. Hence our conclusion was that, even if the approach proposed first in \cite{bagdist1} produces interesting results, in particular, for multiplying  $\delta(x)$ functions, making it possible to define a sort of square of $\delta(x)$, it is not particularly useful when applied to KGFs.  In ordinary QFT the way out to approach this kind of problem is the so-called {\em renormalization theory}, which remove infinities out of diverging results (appearing, e.g., when computing transition probabilities via Feynman graphs) to get finite result. What is really important, and surprising, is that these finite results are often in extremely good agreement with experimental data. However, renormalization appears as an ad hoc procedure, and it would be important to have a theory of elementary particles which do not need any such a procedure, since it produces results which are already finite. As we will see, what we discuss in this paper is indeed a first step in this direction. Hopefully not the last. 

In the past few decades people started to be interested in a sort of {\em extended} version of quantum mechanics, in which it is not required that the observables of the system under analysis, and the Hamiltonian in particular, are self-adjoint. Since the first paper in  1998, \cite{benbot}, this line of research became very famous and several researchers started to work on that, both for its physical implications but also for the interesting mathematics arising when working with observables which are not necessarily self-adjoint. We refer to \cite{benbook}-\cite{bagspringer}, and to references therein, for some books and edited books on several aspects on non-Hermitian\footnote{In this paper Hermitian and self-adjoint will be used as synonymous.} operators. In particular, in the past years we have analyzed in many details some {\em deformed} versions of the canonical (anti-)commutation relations (CCR and CAR), and of other, less knows, algebraic rules, in this new context. Ladder operators of several kind, where the creation operator is not the adjoint of the lowering operator, have been extensively studied and a recent review can be found in \cite{bagspringer}. Among all these ladders, the pseudo-bosonic operators (those which we used to extend the CCR), will be adopted here to define a sort of deformed KGF (our PBKGF) and we will show that interesting features appear. In particular, the two points function of the field will return a finite result, under certain conditions. In view of what discussed before, this result can be seen as a completely different way to {\em regularize} the original KGF, and of course suggests to consider several related problems, i.e. the possibility of bypassing the renormalization in QFT, a very difficult task.

It is surely worth stressing that ours here is not the first attempt to use techniques and results arising for non Hermitian Hamiltonians in a QFT settings. We refer to \cite{benqft1,benqft2,oleg} for some results in this direction. However, up to today, this topic does not seem to us {\em particularly popular}. On the other hand, also in view of the results we are going to discuss in this paper, we believe that a deeper work on this kind of QFT is worth.

The paper is organized as follows:

In the next section we give a short review on pseudo-bosons (PBs), since they will be used as the main tool in our approach.

In Section \ref{sect3} we introduce the PBKGF in 1+1 dimensions,  we study their properties and we show that a two-points function of this field is indeed finite, even in the limit $y\rightarrow x$.

Section \ref{sectconcl} contains our final considerations and plans for the future.

\section{A short review on pseudo-bosons}\label{sect2}

In this section, to keep the paper self-contained, we briefly review few facts on $\D$-pseudo bosons. We refer to \cite{bagspringer} and to \cite{baginbagbook} for more details.

Let $\Hil$ be a given Hilbert space with scalar product $\left<.,.\right>$ and related norm $\|.\|$. Let $a$ and $b$ be two operators
on $\Hil$, with domains $D(a)\subset \Hil$ and $D(b)\subset \Hil$ respectively, $a^\dagger$ and $b^\dagger$ their adjoint, and  $\D$ be a dense subspace of $\Hil$
such that $a^\sharp\D\subseteq\D$ and $b^\sharp\D\subseteq\D$, where with $x^\sharp$ we indicate $x$ or $x^\dagger$. Of course, $\D\subseteq D(a^\sharp)$
and $\D\subseteq D(b^\sharp)$.

\begin{defn}\label{def21}
	The operators $(a,b)$ are $\D$-pseudo bosonic  if, for all $f\in\D$, we have
	\be
	a\,b\,f-b\,a\,f=f.
	\label{A1}\en
\end{defn}

When $b=a^\dagger$, this is simply the canonical commutation relation (CCR) for ordinary bosons. The interesting situation is $b\neq a^\dagger$. Now we consider the following:

\vspace{2mm}

{\bf Assumption $\D$-pb 1.--}  there exists a non-zero $\varphi_{ 0}\in\D$ such that $a\,\varphi_{ 0}=0$.

\vspace{1mm}

{\bf Assumption $\D$-pb 2.--}  there exists a non-zero $\Psi_{ 0}\in\D$ such that $b^\dagger\,\Psi_{ 0}=0$.

\vspace{2mm}

It is obvious that, since $\D$ is stable under the action of $b$ and $a^\dagger$, in particular,  $\varphi_0\in D^\infty(b):=\cap_{k\geq0}D(b^k)$ and  $\Psi_0\in D^\infty(a^\dagger)$, so
that the vectors \be \varphi_n:=\frac{1}{\sqrt{n!}}\,b^n\varphi_0,\qquad \Psi_n:=\frac{1}{\sqrt{n!}}\,{a^\dagger}^n\Psi_0, \label{A2}\en
$n\geq0$, can be defined and they all belong to $\D$, and, as such, to the domains of $a^\sharp$, $b^\sharp$ and $N^\sharp$, where $N=ba$.  Let's put $\F_\Psi=\{\Psi_{ n}, \,n\geq0\}$ and
$\F_\varphi=\{\varphi_{ n}, \,n\geq0\}$.
It is  simple to deduce the following lowering and raising relations:
\be
\left\{
\begin{array}{ll}
	b\,\varphi_n=\sqrt{n+1}\varphi_{n+1}, \qquad\qquad\quad\,\, n\geq 0,\\
	a\,\varphi_0=0,\quad a\varphi_n=\sqrt{n}\,\varphi_{n-1}, \qquad\,\, n\geq 1,\\
	a^\dagger\Psi_n=\sqrt{n+1}\Psi_{n+1}, \qquad\qquad\quad\, n\geq 0,\\
	b^\dagger\Psi_0=0,\quad b^\dagger\Psi_n=\sqrt{n}\,\Psi_{n-1}, \qquad n\geq 1,\\
\end{array}
\right.
\label{A3}\en as well as the eigenvalue equations $N\varphi_n=n\varphi_n$ and  $N^\dagger\Psi_n=n\Psi_n$, $n\geq0$. In particular, as a consequence
of these two last equations,  if we choose the normalization of $\varphi_0$ and $\Psi_0$ in such a way $\left<\varphi_0,\Psi_0\right>=1$, we deduce that
\be \left<\varphi_n,\Psi_m\right>=\delta_{n,m}, \label{A4}\en
for all $n, m\geq0$. Hence $\F_\Psi$ and $\F_\varphi$ are biorthonormal. It is easy to see that, if $b=a^\dagger$, then $\varphi_n=\Psi_n$, so that biorthonormality is replaced by a simpler orthonormality. Also, the relations in (\ref{A3}) collapse, and only one number operator exists, since in this case $N=N^\dagger$.

The analogy with ordinary bosons suggests us to consider the following:

\vspace{2mm}

{\bf Assumption $\D$-pb 3.--}  $\F_\varphi$ is a basis for $\Hil$.

\vspace{1mm}

This is equivalent to requiring that $\F_\Psi$ is a basis for $\Hil$ as well, \cite{chri}. However, several  physical models show that $\F_\varphi$ is {\bf not} necessarily a basis for $\Hil$, but it is still complete in $\Hil$. For this reason we adopt the following weaker version of  Assumption $\D$-pb 3, \cite{baginbagbook}:

\vspace{2mm}

{\bf Assumption $\D$-pbw 3.--}  For some subspace $\G$ dense in $\Hil$, $\F_\varphi$ and $\F_\Psi$ are $\G$-quasi bases.

\vspace{2mm}
This means that, for all $f$ and $g$ in $\G$,
\be
\left<f,g\right>=\sum_{n\geq0}\left<f,\varphi_n\right>\left<\Psi_n,g\right>=\sum_{n\geq0}\left<f,\Psi_n\right>\left<\varphi_n,g\right>,
\label{A4b}
\en
which can be seen as a weak form of the resolution of the identity, restricted to $\G$.

The families $\F_\varphi$ and $\F_\Psi$ can be used to introduce  two densely defined operators $S_\varphi$ and $S_\Psi$ via their
action respectively on  $\F_\Psi$ and $\F_\varphi$: \be
S_\varphi\Psi_{ n}=\varphi_{ n},\qquad
S_\Psi\varphi_{ n}=\Psi_{ n}, \label{213}\en for all $ n$, which also imply that
$\Psi_{ n}=(S_\Psi\,S_\varphi)\Psi_{ n}$ and
$\varphi_{ n}=(S_\varphi \,S_\Psi)\varphi_{ n}$, again for all
$ n$. Of course, these equalities can be extended to the linear spans of the $\varphi_n$'s, $\Lc_\varphi$, and of the $\Psi_n$'s, $\Lc_\Psi$. This means that, for instance, $S_\Psi\,S_\varphi f=f$ and $S_\varphi\,S_\Psi g=g$ for all $f\in\Lc_\Psi$ and $g\in\Lc_\varphi$. With a little abuse of language we could say that $S_\varphi$ is the inverse of $S_\psi$.  Quite often one writes these operators in a bra-ket form:
\be S_\varphi=\sum_{ n}\,
|\varphi_{ n}><\varphi_{ n}|,\qquad S_\Psi=\sum_{ n}
\,|\Psi_{ n}><\Psi_{ n}|, \label{212}\en
where $\left(|f\left>\right<f|\right)g=\left<f,g\right>f$, for all $f,g\in\Hil$.
These expressions may likely be
only formal, since the series are not necessarily convergent in
the uniform topology, as it happens when the operators $S_\varphi$ and $S_\Psi$ are unbounded. However, this is not the case if $\F_\varphi$ and $\F_\Psi$ are Riesz bases. In this case, we call our $\D$-pseudo bosons {\em regular}, \cite{bagspringer,baginbagbook}.
Another interesting aspect of the operators $S_\varphi$ and $S_\psi$ is that they give rise to the following intertwining relations between $N$ and $N^\dagger$:
\be S_\Psi\,N\,g=N^\dagger S_\Psi\,g \quad \mbox{ and }\quad
N\,S_\varphi\, f=S_\varphi\,N^\dagger\,f, \label{219}\en $f\in\Lc_\psi$ and $g\in\Lc_\varphi$, which are in agreement with the fact that $N$ and $N^\dagger$ have the same eigenvalues (other than related eigenvectors, (\ref{213})). 

\vspace{2mm}

These results can be extended to more modes of PBs, as one does for bosons, \cite{rom}. In this case we will have two families of operators, $\{a_j,b_j,\, j\in \J\}$, $\J$ being a certain set of indices (discrete or continuous), such that $a_j\neq b_k^\dagger$ for all $j,k\in \J$, and
\be 
[a_j,b_k]f=\delta_{j,k}f,
\label{214}\en
$\forall f\in\D$. Here $\D$ is, as before, a dense subspace of the Hilbert space where all these operators act and which is invariant under the action of all the $a_j^\sharp$ and $b_j^\sharp$. Formula in (\ref{214}) is the pseudo-bosonic rule if $\J$ is discrete. In case $\J$ is continuous, which is what we will be interesting for us in Section \ref{sect3}, (\ref{214}) must be replaced with
\be 
[a(k),b(q)]f=\delta(k-q)f,
\label{215}\en
$\forall f\in\D$. In this formula $\delta$ is, of course, the Dirac (rather than the Kronecher) delta.

 Similarly with what we have shown before, these are ladder operators and can be used to construct two biorthogonal families
of eigenstates of $N_j=b_ja_j$ and $N_j^\dagger=a_j^\dagger b_j^\dagger$, $j\in\J$, or of $N(k)=b(k)a(k)$ and of its adjoint in the continuous case. All what has been stated for a single mode of PBs can be extended to many modes without particular problems.

In the next section we will be particularly interested in a specific version of PBs, those related to the Swanson Hamiltonian, \cite{swan}. In \cite{bag2010} we have shown how this Hamiltonian (together with others) can be rewritten in terms of PBs, and we have applied and extended these results in several ways, including also a distributional approach to some quantum mechanical systems, \cite{bagspringer,bag2010,bag2022}. In particular, we have shown that the manifestly non-Hermitian Hamiltonian
$$
H_\theta=\frac{1}{2}\left(p^2+x^2\right)-\frac{i}{2}\,\tan(2\theta)\left(p^2-x^2\right),
$$
proposed in \cite{dapro} and strictly connected to the Swanson Hamiltonian, can be rewritten in terms of pseudo-bosonic operators. Here $\theta\in\left(-\frac{\pi}{4},\frac{\pi}{4}\right)\setminus\{0\}=:I$. We see that $H_\theta^\dagger=H_{-\theta}\neq H_\theta$, for all $\theta\in I$. The operators $x$ and $p$ are the self-adjoint position and momentum, and they satisfy $[x,p]=i\1$. Introducing the bosonic annihilation and creation operators $c=\frac{1}{\sqrt{2}}(x+ip)$ and $c^\dagger=\frac{1}{\sqrt{2}}(x-ip)$, and putting further
\be
\left\{
\begin{array}{ll}
	A_\theta=\cos(\theta)\,c+i\sin(\theta)\,c^\dagger,  \\
	B_\theta=\cos(\theta)\,c^\dagger+i\sin(\theta)\,c,
\end{array}
\right.
\label{216}\en
we have rewritten $H_\theta$ as
\be
H_\theta=\omega_\theta\left(B_\theta\,A_\theta+\frac{1}{2}\1\right).
\label{217}\en
Here $\omega_\theta=\frac{1}{\cos(2\theta)}$, which  is well defined since $\cos(2\theta)\neq0$ for all $\theta\in I$. It is clear that $A_\theta^\dagger\neq B_\theta$ and that $[A_\theta,B_\theta]=\1$. We refer to \cite{bag2010} for a full analysis of this model, including the expression of the eigenstates of $H_\theta$ and $H_\theta^\dagger$. Here we will see how to use (\ref{216}), extended to a continuos set of modes, in the construction of our PBKGF.

  \section{The pseudo-bosonic Klein Gordon field}\label{sect3}
  
As it is well known, the (standard) KGF is the solution of  the following second order differential equation
\be
\left( \frac{\partial^2}{\partial t^2}-\frac{\partial^2}{\partial x^2}+m^2 \right)
\varphi (x,t)=0
\label{31}
\en
which is quantized by assuming the equal time canonical commutation relations
\bea
\left\{
\begin{array}{ll}
	& [\varphi(x,t), \varphi(x',t)]=0 \\
	& [\dot \varphi(x,t), \dot \varphi(x',t)]=0, \\
	& [\varphi(x,t), \dot \varphi(x',t)]=i\delta (x-x').\\
\end{array}
\right.
\label{32}
\ena
Following the notation and the main steps of \cite{bd}, we expand the solution of the
Klein-Gordon equation in plane waves, 
\be
\varphi(x,t) =\int_{-\infty}^{\infty}\frac{dk}{\sqrt{4\pi \omega_k}} \left[ c(k)
e^{ikx-i\omega_kt}+ c^{\dagger}(k) e^{-ikx+i\omega_kt} \right],
\label{33}
\en
where $\omega_k =\sqrt{k^2+m^2}$ and the operators $c(k)$ and its Hermitean conjugate 
$c^{\dagger}(k)$ are the coefficients of the expansion. 
They satisfy the CCRs
\be
[c(k),c(q)]=[c^{\dagger}(k),c^{\dagger}(q)]=0, \quad [c(k),c^{\dagger}(q)]=
\delta(k-q)\,\1,
\label{34}
\en
for all $k,q\in\mathbb{R}$.
More explicitly, assuming (\ref{33}), it is possible to show that (\ref{34}) implies (\ref{32}), and vice-versa. 

\vspace{2mm}

{\bf Remarks:--} (1) The expansion in (\ref{33}) can be thought as a classical function (or distribution), rather than an operator, if we don't explicitly require (\ref{32}) or (\ref{34}), i.e., if we consider $c(k)$ and $c^\dagger(k)$ simply as $k-$dependent (commuting) functions. In this case, of course, $c^\dagger(k)$ should be replaced by $\overline{c(k)}$.

\vspace{1mm}

(2) In this section we will mostly work {\em formally}, i.e. not paying much attention to the fact that the fields we are considering, here and in the rest of the section, are indeed operator valued distributions. Our main effort here is to show that, moving from bosonic to pseudo-bosonic operators in an expansion like that in (\ref{33}), some infinite physical quantity turns out to become finite. We will say more on this aspect in Section \ref{sectconcl}.

\vspace{2mm}

Let us call $e_0$ the ground state of the theory, \cite{bd}. This is defined by
requiring that  $c(k) e_0 =0$,  $\forall k\in \mathbb{R}$. 

Interesting quantities to compute are the
expectation values in $e_0$ of the field $\varphi(x,t)$ and of the product of the
field, $\varphi(x,t)\varphi(x',t')$. It is well known that problems arise when we try to compute, in particular, the mean value of the product
$\varphi(x,t)\varphi(x,t)$ on $e_0$. In fact, while the one-point function of the field is well defined, and trivial, $
\langle e_0, \varphi(x,t) e_0\rangle  =0,
$
we also find that
\be
\Delta_+(x,y;t,s)= \langle e_0, \varphi(x,t) \varphi(y,s) e_0\rangle =\int_{-\infty}^{\infty}
\frac{dk}{4\pi \omega_k} e^{ik (x-y)-i\omega_k(t-s)},
\label{35}
\en
which, in the limit $(y,s) \rightarrow (x,t)$, diverges logaritmically\footnote{Incidentally we recall
 that in four dimensions the analogous divergence is
quadratic.}, \cite{bd,das}. This is just the first divergence of many others which one has to face with when working with QFT.

\vspace{2mm}

Let us now consider the following extended version of (\ref{33}):
\be
\Phi_\theta(x,t) =\int_{-\infty}^{\infty}\frac{dk}{\sqrt{4\pi \omega_k}} \left[ A_\theta(k)
e^{ikx-i\omega_kt}+ B_\theta(k) e^{-ikx+i\omega_kt} \right],
\label{36}
\en
where $\theta$ is a (real) parameter in $I$. For the moment we assume that $A_\theta(k)$ and $B_\theta(k)$ are simply coefficients of the expansion of $\Phi_\theta(x,t)$ in plane waves, and that, in principle, $B_\theta(k)$ is not necessarily the complex conjugate of $A_\theta(k)$. In other words, both $A_\theta(k)$ and $B_\theta(k)$ are functions and $A_\theta(k)\neq\overline{B_\theta(k)}$, in principle.

 This is similar to what happens for charged fields, which are not Hermitian, see \cite{das} for instance. Hence this suggests to introduce the following Lagrangian density for our field:
\be
\Lc_\theta=\dot \Phi_\theta^\dagger\,\dot\Phi_\theta-(\partial_x\Phi_\theta)^\dagger(\partial_x\Phi_\theta)-m^2\Phi_\theta^\dagger\Phi_\theta,
\label{39}\en
where, to simplify the notation, we have omitted everywhere the dependence on $x$ and $t$. We stress once more that, at this stage, we are looking at the fields appearing in (\ref{39}) as classical objects and, for this reason, the $\dagger$ symbol should be understood as a simple complex conjugation. We prefer to keep this symbol here since it will be relevant soon, when quantizing the field. Replacing $\Lc_\theta$ in the Euler-Lagrange equations
$$
\frac{d}{dt}\left(\frac{\partial \Lc_\theta}{\partial\dot\Phi_\theta^\dagger}\right)+\frac{\partial}{\partial x}\left(\frac{\partial \Lc_\theta}{\partial(\partial_x\Phi_\theta)^\dagger}\right)-\frac{\partial\Lc_\theta}{\partial\Phi_\theta^\dagger}=0,
$$
and its adjoint, we recover
\be\left(\partial_t^2-\partial_x^2+m^2\right)\Phi_\theta(x,t)=\left(\partial_t^2-\partial_x^2+m^2\right)\Phi_\theta^\dagger(x,t)=0,
\label{310}\en
as expected. From (\ref{39}) we also compute the conjugate momenta of $\Phi_\theta(x,t)$ and $\Phi_\theta^\dagger(x,t)$:
\be
\Pi_\theta(x,t)=\frac{\partial \Lc_\theta}{\partial\dot\Phi_\theta^\dagger}=\dot\Phi_\theta(x,t), \qquad \Pi_\theta^\dagger(x,t)=\frac{\partial \Lc_\theta}{\partial\dot\Phi_\theta}=\dot\Phi_\theta^\dagger(x,t).
\label{311}\en
Then the Hamiltonian density is
\be
\HH_\theta=\Pi_\theta\dot\Phi_\theta^\dagger+\Pi_\theta^\dagger\dot\Phi_\theta-\Lc_\theta=\Pi_\theta^\dagger\Pi_\theta+(\partial_x\Phi_\theta)^\dagger(\partial_x\Phi_\theta)+m^2\Phi_\theta^\dagger\Phi_\theta.
\label{312}\en

To quantize the system we now assume that $A_\theta(k)$ and $B_\theta(k)$ is a continuous family of pseudo-bosonic operators obeying the following commutation relations, which extend those in (\ref{34}):
\be
[A_\theta(k),A_\theta(q)]=[B_\theta(k),B_\theta(q)]=0, \quad [A_\theta(k),B_\theta(q)]=
\delta(k-q)\,\1,
\label{37}
\en
for all $k,q\in\mathbb{R}$. For concreteness' sake (and because our choice will produce interesting results) we will consider here a specific class of operators $A_\theta(k)$ and $B_\theta(k)$. We know, see \cite{bagspringer}, that many other choices are also possible, but we will not consider these alternatives here. To be concrete, we will assume that $A_\theta(k)$ and $B_\theta(k)$ can be written in terms of $c(k)$ and $c^\dagger(k)$ satisfying (\ref{34}) as follows:
\be
\left\{
\begin{array}{ll}
	A_\theta(k)=\cos(\theta)\,c(k)+i\sin(\theta)\,c^\dagger(k),  \\
	B_\theta(k)=\cos(\theta)\,c^\dagger(k)+i\sin(\theta)\,c(k),
\end{array}
\right.
\label{38}\en
$\theta\in I$ and $k\in\mathbb{R}$, which is a multi-mode version of (\ref{216}). It is clear that $\Phi_0(x,t)=\varphi(x,t)$ since, when $\theta=0$, $A_0(k)=c(k)$ and $B_0(k)=c^\dagger(k)$.

The first obvious remark is that, since for all $\theta\in I\setminus\{0\}$ $B_\theta(k)^\dagger\neq A_\theta(k)$, $(\Phi_\theta(x,t))^\dagger\neq \Phi_\theta(x,t)$, while $(\varphi(x,t))^\dagger=\varphi(x,t)$.

Formulas (\ref{37}) are now deduced if we assume the following equal time canonical commutation rules between the fields and their conjugate momenta
\be
[\Phi_\theta(x,t),\Phi_\theta(y,t)]=[\Pi_\theta(x,t),\Pi_\theta(y,t)]=0, \qquad [\Phi_\theta(x,t),\Pi_\theta(y,t)]=i\delta(x-y),
\label{313}\en
with similar rules for $\Phi_\theta^\dagger(x,t)$ and $\Pi_\theta^\dagger(x,t)$:
\be
[\Phi_\theta^\dagger(x,t),\Phi_\theta^\dagger(y,t)]=[\Pi_\theta^\dagger(x,t),\Pi_\theta^\dagger(y,t)]=0, \qquad [\Phi_\theta^\dagger(x,t),\Pi_\theta^\dagger(y,t)]=i\delta(x-y),
\label{314}\en

If we now insert (\ref{36}) and
\be
\Pi_\theta(x,t)=\dot\Phi_\theta(x,t)=-i\int_{-\infty}^{\infty}\sqrt{\frac{\omega_k}{4\pi}} \left[ A_\theta(k)
e^{ikx-i\omega_kt}- B_\theta(k) e^{-ikx+i\omega_kt} \right]dk,
\label{316}\en
together with their adjoints, in formula (\ref{312}) for $\HH_\theta$, and then we integrate over $x$, we find the following expression for the Hamiltonian operator
\be
H_\theta=\int_{\mathbb{R}}\HH_\theta(x,t)\,dx=\int_{\mathbb{R}}dk\,\omega_k\left(B_\theta^\dagger(k)B_\theta(k)+A_\theta^\dagger(k)A_\theta(k)\right)
\label{317}\en
which is (formally) self-adjoint, and which reduces to the usual form of the Hamiltonian of the KGF if $\theta=0$. In fact, there is more than this: in terms of the $c(k)$ and $c^\dagger(k)$'s it turns out that
\be
H_\theta=\int_{\mathbb{R}}dk\,\omega_k\left(c(k)c^\dagger(k)+c^\dagger(k)c(k)\right),
\label{318}\en
which is exactly the Hamiltonian of KGF. Hence, in particular, $H_\theta$ does not depend on $\theta$, contrarily to  $\HH_\theta$. Notice that $H_\theta$ also does not depend explicitly on time, as it is clear from its integral expression in terms of pseudo-bosonic or of bosonic operators. 

Summarizing what we have found so far is not particularly different from what is known for a standard (complex) KGF. Another similar result is the following: we can recover $A_\theta(k)$ and $B_\theta(k)$ from $\Phi_\theta(x,t)$ and $\Pi_\theta(x,t)$ as follows:
\be
A_\theta(k)=\frac{1}{4\pi\omega_k}\int_{\mathbb{R}}dxe^{-ikx+i\omega_k t}\left(\omega_k\Phi_\theta(x,t)+i\Pi_\theta(x,t)\right),
\label{319}\en
and
\be
B_\theta(k)=\frac{1}{4\pi\omega_k}\int_{\mathbb{R}}dxe^{ikx-i\omega_k t}\left(\omega_k\Phi_\theta(x,t)-i\Pi_\theta(x,t)\right),
\label{320}\en
this latter being clearly different from $A_\theta^\dagger(k)$.
These are essentially the same inverse formulas holding true for the KGF, and can be deduced with similar computations. 

It is interesting to observe that, because of (\ref{38}), we have
\be
B_\theta^\dagger(k)=A_{-\theta}(k), \qquad A_\theta^\dagger(k)=B_{-\theta}(k),
\label{321}\en
and, equivalently
\be
\Phi_\theta^\dagger(x,t)=\Phi_{-\theta}(x,t), \qquad \Pi_\theta^\dagger(x,t)=\Pi_{-\theta}(x,t).
\label{322}\en
For this reason, in particular, we could rewrite $H_\theta$ in (\ref{316}) replacing $B_\theta^\dagger(k)$ with $A_{-\theta}(k)$ and $A_\theta^\dagger(k)$ with $B_{-\theta}(k)$.

\vspace{2mm}

{\bf Remark:--} Other commutation rules can be computed using (\ref{36}), (\ref{38}) and  (\ref{316}). In particular we get
\be
[A_\theta(k),A_{-\theta}(q)]=-[B_\theta(k),B_{-\theta}(q)]=-i\sin2\theta\delta(k-k)\1,
\label{comm1}\en
\be
[\Phi_\theta(x,t),\Phi_\theta^\dagger(y,t)]=-\frac{\sin2\theta}{2\pi}\int\frac{dk}{\omega_k}e^{ik(x-y)}\sin(2\omega_kt)\,\1,
\label{comm2}\en
and
\be
[\Pi_\theta(x,t),\Pi_\theta^\dagger(y,t)]=\frac{\sin2\theta}{2\pi}\int\,dk\,\omega_k e^{ik(x-y)}\sin(2\omega_kt)\,\1,
\label{comm3}\en
which all return {\em standard} results (i.e., those we know for ordinary KGF) if $\theta=0$. In case $\sin2\theta\neq0$ we still observe that $[\Phi_\theta(x,0),\Phi_\theta^\dagger(y,0)]=[\Phi_\theta(x,0),\Phi_\theta^\dagger(y,0)]=0$, while these commutators are different from zero for $t>0$, in general.

\subsection{A finite two points function}

For what we have to do here, it is convenient to rewrite now $\Phi_\theta(x,t)$ in (\ref{36}) in terms of the bosonic operators in (\ref{34}). We get
\be
\Phi_\theta(x,t) =\int_{-\infty}^{\infty}\frac{dk}{\sqrt{4\pi \omega_k}} \left[ c(k)\alpha_\theta(k;x,t)+c^\dagger(k)\beta_\theta(k;x,t)
 \right],
\label{323}
\en
where
\be
\alpha_\theta(k;x,t)=\cos\theta e^{ikx-i\omega_kt}+ i\,\sin\theta e^{-ikx+i\omega_kt}, \qquad 
\beta_\theta(k;x,t)=\cos\theta e^{-ikx+i\omega_kt}+ i\,\sin\theta e^{ikx-i\omega_kt}.
\label{324}\en
Using this expression for $\Phi_{\theta}(x,t)$ it is clear that
\be
F_\theta^{(1)}(x,t)=\langle e_0,\Phi_\theta(x,t)e_0\rangle=0,
\label{325}\en
for all $x$ and $t$. Similarly we have
\be
G_\theta^{(1)}(x,t)=\langle e_0,\Pi_\theta(x,t)e_0\rangle=0,
\label{326}\en
for all $x$ and $t$. Here we have used the following expression for $\Pi_\theta(x,t)$:
\be \Pi_\theta(x,t)=-i\int_{-\infty}^{\infty}\sqrt{\frac{\omega_k}{4\pi}}\,dk \left[ c(k)\alpha_{-\theta}(k;x,t)-c^\dagger(k)\beta_{-\theta}(k;x,t)
	\right],\label{327}\en
using the same definitions given in (\ref{324}).

Let us now consider the two-points function $F_\theta^{(2)}(x,t;y,s)=\langle e_0,\Phi_\theta(x,t)\Phi_\theta(y,s)e_0\rangle$. According to what we have shown in (\ref{35}), see also \cite{bd,das}, what is {\em dangerous} is the limit of this function when $(y,s)\rightarrow(x,t)$. For this reason we will only consider here $F_\theta^{(2)}(x,0;y,0)$, and then discuss what happens when $y\rightarrow x$. We will find a different behavior for $x=0$ and for $x\neq0$.

Using (\ref{323}), together with $c(k)e_0=0$, it is easy to find that
\be
F_\theta^{(2)}(x,0;y,0)=\int_{-\infty}^{\infty}\frac{dk}{{4\pi \omega_k}}\overline{\beta_{-\theta}(k;x,0)}\,\beta_{\theta}(k;y,0),
\label{328}\en
which, when $y\rightarrow x$, produces
\be
F_\theta^{(2)}(x,0;x,0)=\cos(2\theta)\int_{-\infty}^{\infty}\frac{dk}{{4\pi \omega_k}}+i\sin(2\theta)\int_{-\infty}^{\infty}\frac{dk}{{4\pi \omega_k}}e^{2ikx}.
\label{329}\en
Our first remark is the following: if $x=0$ we get $F_\theta^{(2)}(0,0;0,0)=e^{2i\theta}\int_{-\infty}^{\infty}\frac{dk}{{4\pi \omega_k}}$, and we are back to what we have seen in (\ref{35}) and after, for the standard KGF: $F_\theta^{(2)}(0,0;0,0)=\infty$. However, if $x\neq0$, the situation is quite different. In this case, and if we further fix\footnote{This is not the only useful choice of $\theta$, as we will see. Other choices are also possible.} $\theta=\frac{\pi}{4}$, we deduce that
\be
F_{\frac{\pi}{4}}^{(2)}(x,0;x,0)=i\,\int_{-\infty}^{\infty}\frac{dk}{{4\pi \omega_k}}e^{2ikx}=\frac{i}{2\pi}\int_0^\infty\frac{dk}{{\omega_k}}\cos(2kx),
\label{330}\en
with easy computations. This integral can be written in terms of the Bessel function $K_0(x)$ as follows:
\be
F_{\frac{\pi}{4}}^{(2)}(x,0;x,0)=\frac{i}{2\pi}K_0(2m|x|).
\label{331}\en
An interesting remark is that $K_0(2m|x|)$ diverges when $x\rightarrow 0$, in agreement with what we have found before for $F_\theta^{(2)}(0,0;0,0)$, but, if $x$ is not zero, we get a finite result. This is different from what we have found in \cite{bagdist2}, and from what happens for ordinary KGF, where $F_\theta^{(2)}(x,0;x,0)$ diverges also for $x\neq0$. In other words, with the particular choice of $\theta=\frac{\pi}{4}$ (but also taking, e.g., $\theta=\frac{3\pi}{4}$) the two-points function $F_\theta^{(2)}(x,0;x,0)$ turns out to be finite for all $x\neq0$. 

Similarly, we can consider
\be
G_\theta^{(2)}(x,t;y,s)=\langle e_0,\Pi_\theta(x,t)\Pi_\theta(y,s)e_0\rangle,
\label{332}\en
and check if also $G_\theta^{(2)}(x,0;x,0)$ is finite for all $x\neq 0$. However, this seems a bit more complicated. In fact, in analogy with (\ref{330}), we find that
\be
G_{\frac{\pi}{4}}^{(2)}(x,0;x,0)=-\frac{i}{2\pi}\int_0^\infty\omega_k\cos(2kx)dk,
\label{333}\en
which , if $x=0$ diverges quadratically in $k$. If $x\neq0$ the situation is slightly more involved. Despite of its apparent divergence, it is still possible to understand that $G_{\frac{\pi}{4}}^{(2)}(x,0;x,0)$ is an {\em manageable quantity}. For that, it is convenient to rewrite it as
\be
G_{\frac{\pi}{4}}^{(2)}(x,0;x,0)=-\frac{i}{4\pi}\int_{\mathbb{R}}\omega_ke^{2ikx}\,dk,
\label{334}\en
and then to consider its convolution with a (generic) function $f(x)\in\scr$, the set of test functions, \cite{kor}:
\be
G^{(2)}[f](y)=\int_{\mathbb{R}}f(y-x)G_{\frac{\pi}{4}}^{(2)}(x,0;x,0)dx=-\frac{i}{4\pi}\int_{\mathbb{R}}\omega_k\left(\int_{\mathbb{R}}f(y-x)e^{2ikx}dx\right)dk,
\label{335}\en
with a change of integration. Hence we get 
\be
G^{(2)}[f](y)=-\frac{i}{2\sqrt{2\pi}}\int_{\mathbb{R}}\omega_k \hat f(2k)e^{2iky}\,dk,
\label{336}\en
where $\hat f(2k)$ is the Fourier transform of $f(x)$ computed in $2k$. It is clear that $G^{(2)}[f](y)$ is well defined for all possible $y$ and for all $f(x)\in\scr$, since $\hat f\in\scr$ as well. This means that $\hat f(2k)$ goes to zero very fast, while $\omega_k$ diverges linearly in $|k|$: their product is integrable. Furthermore, it is also possible to (try to) recover $G_{\frac{\pi}{4}}^{(2)}(x,0;x,0)$ out of $G^{(2)}[f](y)$. In fact, this could be formally done  by simply fixing $f(x)=\delta(x)$, but this is not in agreement with our request to have $f(x)\in\scr$. However, see \cite{kor,bagdist1,bagdist2,bagdist3}, we could consider a $\delta$-sequence of functions in $\D(\mathbb{R})$, and therefore of $\scr$. Some useful properties, and the construction of these sequences, are listed in the following theorem, see \cite{bagdist1} for instance:

\begin{thm}
	Let $\phi \in {\cal D}({\bf R})$ be a given function with supp $\phi \subseteq
	[-1,1]$ and $\int \phi (x) \, dx =1$. We call $\delta-$sequence the sequence
	$\delta_n,\, n\in {\bf N},$ defined by $\delta_n(x) \equiv n\, \phi(nx)$.
	
	Then, $\forall \,  T \in {\cal D'}({\bf R})$, the set of distributions, the convolution $T_n \equiv
	T*\delta_n$ is a $C^{\infty}-$function, for any fixed $n\in {\bf N}$. This
	sequence converges to $T$ in the topology of ${\cal D'}$, when $n \rightarrow
	\infty$.
	
	Moreover, if $T(x)$ is a continuous function with compact support, then $T_n$
	converges uniformly to $T(x)$.
\end{thm}

In other words, even if we cannot take $f(x)=\delta(x)$ in (\ref{335}), we could still replace $f(x)$ with a sequence of functions $\delta_n(x)$ in $\scr$ converging to $\delta(x)$ as in this theorem, and then look for the  limit of $G^{(2)}[\delta_n](y)$ when $n\rightarrow\infty$. This distributional analysis of $G^{(2)}_\theta$ is work in progress.

\vspace{1mm}

As a final remark we observe that for $\theta=\frac{\pi}{4}$ our pseudo-bosonic operators can be written as
\be
A_{\frac{\pi}{4}}(k)=\frac{1+i}{2}(x(k)+p(k)), \qquad B_{\frac{\pi}{4}}(k)=\frac{1+i}{2}(x(k)-p(k)),
\label{338}\en
where $x(k)=\frac{1}{\sqrt{2}}(c(k)+c^\dagger(k))$ and $p(k)=\frac{1}{\sqrt{2}\,i}(c(k)-c^\dagger(k))$ can be seen as a sort of {\em multi-mode} version of the standard position and momentum operators. Formula (\ref{338}) clarifies once more that $A_{\frac{\pi}{4}}(k)$ is not the adjoint of $B_{\frac{\pi}{4}}(k)$. Also, a straight computation confirms that they obey the pseudo-bosonic rule $[A_{\frac{\pi}{4}}(k),B_{\frac{\pi}{4}}(q)]=\delta(k-q)\1$, as they should.

\section{Conclusions}\label{sectconcl}

This paper is a first {\em incursion} of PBs in QFT and, in our opinion, the results we have deduced are rather promising and suggest to carry on the analysis in many ways. In particular, we should consider other possible divergences arising in other relevant $n$-points functions and see if they are cancelled, or controlled, replacing bosons with PBs. The same should be done when considering higher spatial dimensions. Also, the KGF we have discussed here is a free field. Interacting fields make the system much more complicated, and more and more diverging Feynman graphs arise. Hence it is natural to check if PBs can be useful also for these more complicated systems. 

We should also remind that the specific form of the pseudo-bosonic ladders we have considered here, see (\ref{38}), are those arising from the Swanson model. But many other PBs also exist, \cite{bagspringer}. An analysis of these alternative operators in this context would be interesting. Together with the use of, e.g., pseudo-fermions, \cite{bagspringer}, in connection with the Dirac field or other fermionic quantum fields. This analysis could possibly be relevant in connection with what is discussed in \cite{benqft1,benqft2,oleg}.

We end this paper with a final remark concerning the apparent (lack of) mathematical rigour in some parts of this paper. This aspect was not so relevant for us {\bf here}, since, as we have already pointed out, we were more interested in understanding if PBs can somehow contribute to a very old problem of QFT, that of diverging quantities. Our results here suggest that this is the case. Hence it makes sense to try to put our results on more rigorous bases, and this is part of our future plans.

\section*{Acknowledgements}

 The author acknowledges partial financial support from Palermo University and from G.N.F.M. of the INdAM. This work has also been partially supported by the PRIN grant {\em Transport phenomena in low dimensional
 		structures: models, simulations and theoretical aspects}- project code 2022TMW2PY - CUP B53D23009500006, and partially by  project ICON-Q, Partenariato Esteso NQSTI - PE00000023, Spoke 2.

\end{document}